\input{aipcheck}

\documentclass[
    ,final            
  ]
  {aipproc}

\layoutstyle{8x11single}

\def\Meszaros{M\'esz\'aros~}

\def\cm3{\mbox{cm}^{-3}}

\def\tdec2{t_{dec,2}}
\def\cm{\mbox{cm}}

\begin{document}

\title{ Non-thermal emission from supernova shock breakout and the
origin of the X-ray transient associated with SN2008D}


\author{Xiang-Yu  Wang}{
  address={Department of Astronomy, Nanjing University,
Nanjing 210093, China }}

\author{Zhuo Li}{
  address={Department of Astronomy, Peking University, Beijing 100871, China} }
\author{Eli Waxman}{
  address={Physics Faculty, Weizmann Institute of Science, Rehovot
76100, Israel }}
\author{P. M\'esz\'aros}{
  address={Department of Astronomy and
Astrophysics, Penn. State University, University Park, PA 16802,
USA} }

\begin{abstract}
We suggest that non-thermal emission can be produced by multiple
scatterings of the photons between the supernova ejecta and
pre-shock material in supernova shock breakout. Such
bulk-Comptonization process may significantly change the original
thermal photon spectrum, forming a power-law non-thermal component
at higher energies. We then show that the luminous X-ray outburst
XRO081009 associated with SN2008D is likely to be such shock
breakout emission from an ordinary type Ib/c supernova.

\end{abstract}


\maketitle

\section{Introduction}
Supernova shock breakout has been predicted for a few decades
(e.g. \cite{Colgate 1974}; \cite{Klein & Chevalier 1978};
\cite{Ensman & Burrows 1992}; \cite{Matzner & Mckee 1999}). In
core collapse supernovae, a shock wave is generated which
propagates through the progenitor star and ejects the envelope. As
the shock propagates through the envelope, it is mediated by
radiation: the post shock energy density is dominated by
radiation, and the shock transition is mediated by Compton
scattering. As the shock approaches the edge of the star, the
optical depth of the plasma lying ahead of the shock decreases. At
the radius where the optical depth drops to $\sim c/v_s$, where
$v_s$ is the shock velocity, the shock wave emerges through the
surface and the radiation decouples, accompanied by a very bright
ultraviolet/X-ray burst of radiation. The radiation spectrum is
expected to be thermal-dominated.  The term "shock breakout" is
commonly used to refer to the emergence of the shock from the edge
of the star. However, if the star is surrounded by an optically
thick wind, the radiation mediated shock would continue to
propagate into the wind, up to the point where the wind optical
depth drops below $\sim c/v_s$. In this case, shock breakout,
occurs as the shock propagates through the wind, at a radius which
may be significantly larger than the star's radius. We use here
the term "shock breakout" to denote this transition in general,
regardless of whether it occurs as the shock reaches the edge of
the star or further out within an optically thick wind.

This kind of radiation has previously never been directly detected
from any normal supernovae due to its transient nature and its
very early  (minutes to hours) occurrence, in the absence of a
suitably prompt trigger alert. On 2006 February 18, thanks to its
sensitive gamma-ray trigger and rapid slewing capability, {\it
Swift} has detected early thermal X-rays emission from a supernova
(SN2006aj) associated with a low-luminosity GRB, namely GRB060218
(e.g. \cite{Campana et al. 2006}), which has been interpreted as
arising from the breakout of a radiation-dominated shock
(\cite{Waxman et al. 2007}; \cite{Campana et al. 2006}). The long
duration ($\sim 3000$ s) of the thermal X-ray emission of
GRB060218/SN2006aj suggests that the shock breaks out from a
dense, optically-thick wind surrounding the progenitor
star{\footnote{Another possibility is that shock breaks out from
an optically thick shell pre-ejected from the progenitor (the
required shell mass is only $10^{-7} M_\odot$, see \cite{Campana
et al. 2006}).}}.  The energetic of the thermal and non-thermal
emission is of the order of $10^{49} {\rm ergs}$, which is much
larger than the predicted radiation energy from a normal type Ib/c
supernova (\cite{Matzner & Mckee 1999}). Such a large energy
release could be due to a larger kinetic energy in hypernova
SN2006aj and/or an central engine that drives a jet which is,
however, chocked in the outward propagation.

On 2008 January 9, a bright X-ray outburst XRO080109 was
serendipitously discovered during a scheduled {\em Swift}
observations of the galaxy NGC 2770 (\cite{Soderberg et al.
2008}). An ordinary type Ib/c supernova in coincident with this
outburst was later spectroscopically identified and named SN2008D.
This outburst has an energy of $2\times10^{46} {\rm erg}$, which
is three order of magnitudes smaller than even low-luminosity
GRBs, but is astonishingly close to the predicted shock breakout
radiation energy from a normal type Ib/c supernova (\cite{Matzner
& Mckee 1999}).

The most mysterious thing of this X-ray outburst is the
non-thermal spectrum, which is in  contrast with the thermal
spectrum predicted by the shock break theory. There is a general
agreement that a power-law spectrum (with photon index
$\Gamma=2.3\pm0.3$) provides a better fit for this x-ray outburst
than a blackbody (\cite{Soderberg et al. 2008}; \cite{Xu et al.
2008}; \cite{Li 2008}; \cite{Mazzali et al. 2008}).  The temporal
evolution is characterized by a fast rise and exponential decay,
with a FWHM duration about 100 s. The rise time is fitted to be
$63\pm 7$ s (\cite{Soderberg et al. 2008}).

\section{Non-thermal emission from supernova shock breakout due to
bulk motion Comptonization }

Blandford \& Payne\cite{Blandford & Payne 1981} first noted the
importance of bulk motion acceleration of photons in a
radiation-dominated shock. They found that photons are
preferentially upscattered by the bulk motion rather than by the
thermal motions of the electrons, and a power-law spectrum
extending to high energies forms, when the electron thermal
velocity is less than the shock velocity $v_s$. Repeated
scatterings using the energy of the bulk motions of two
approaching relativistic shells in the context of GRB internal
shocks was studied by \cite{Gruzinov & Meszaros 2000}. They found
that the seed synchrotron photons can be boosted to much higher
energies, which is confirmed by their Monte Carlo simulations.
This process is equivalent to the Fermi acceleration mechanism of
particles   or  photon scattering off Alvf\'{e}n waves
(\cite{Thompson 1994}), but here the mechanism, instead, uses the
relative bulk motion and accelerates photons.

Let's consider a mildly relativistic ejecta driving a
radiation-dominated shock into the stellar envelope of supernova
progenitor or into an optically thick wind (or pre-ejected shell)
surrounding it. This mildly relativistic ejecta could result from
the shock acceleration in the surface of the type Ib/c supernova
progenitor (\cite{Matzner & Mckee 1999}) or from a  jet that is
chocked and expands sideways in the outward propagation inside the
star. Once the optical depth of the material in front of the shock
drops below $c/v_s$ (where $v_s$ is the shock velocity), the
photons escape and produce a breakout flash. Since the Thompson
scattering optical depth is non-negligible in front of the shock
while the shock is breaking out, { some fraction of the thermal
photons will be scattered back. The back-scattered photons will be
scattered forward by the expanding ejecta or shocked plasma,
boosting up their energy. The backward-forward scattering cycle
may repeat itself many times for some fraction of the photons,
boosting their energy by a large factor} \cite{Wang et al. 2007}.

\subsection{A qualitative description}
{ Ignoring the time dependence of $\tau$, the optical depth ahead
of the shock, the physical situation is  similar to that of
Comptonization by a thermal electron plasma. In many cases, the
electrons are cold and their momentum is dominated by the bulk
motion.} Assuming each scattering amplifies the photon energy by a
factor $A$, the energy of a photon escaping after $k$ scatterings
is $\varepsilon_k=\varepsilon_i A^k$, where $\varepsilon_i$ and
$\varepsilon_k$ are the initial and final photon energies
respectively. { For fixed $\tau$,} a photon scattered by the
ejecta has a probability $1-e^{-\tau}$ to be scattered back
towards the ejecta, and a probability $e^{-\tau}$ to escape. The
probability for a photon to undergo $k$ scatterings before
escaping is $(1-e^{-\tau})^k$, and since the photon energy is
multiplied by $A$ per scattering, the escaping photon intensity
would have a power-law shape

\begin{equation}
F(\varepsilon_k)\sim F(\varepsilon_i)(1-e^{-\tau})^k\sim
F(\varepsilon_i)(\varepsilon_k/\varepsilon_i)^{-\alpha}
\end{equation}
with
\begin{equation}
\alpha=-{\rm ln}(1-e^{-\tau})/{\rm ln}A.
\end{equation}
The  photon energy amplification factor $A$ is determined by the
kinetic energy of the electrons. For trans-relativistic electrons
{ and isotropic photon distributions,
$A\sim\Gamma^2(1+\beta^2/3)$}. The power-law spectrum extends to a
cutoff energy, { which is the smaller of the electron kinetic
energy, $\sim(\Gamma-1)m_e c^2$, and its rest mass $m_ec^2$ (due
to the Klein-Nishina effect). }

An important difference between the usual thermal electron (or
bulk) Comptonization case and the current case is that in the
present case the scattering optical depth decreases with time as
the mildly relativistic ejecta moves outward. Initially, when the
optical depth $\tau\geq1$, the slope of $\nu F_\nu$ is positive
and most radiation is emitted at high energies. As $\tau$
decreases, the spectrum becomes softer and softer, and at late
times the spectrum is composed of a thermal peak plus a weak
high-energy power-law tail. In general, we expect a noticeable
spectral softening of the nonthermal emission with time. This
spectral softening is expected to be accompanied by a decrease in
the Compton luminosity. At early time, when the effective Compton
parameter $Y=A(1-e^{-\tau})>1$, the Compton luminosity may exceed
the thermal luminosity (it is limited by the kinetic energy of the
ejecta $E_k$). We expect the Compton luminosity to decrease with
time, as $Y$ decreases.

The x-ray or gamma-ray light curves produced in this model
generally have a simple profile without multi-peak structure. The
characteristic variability timescale $\delta t$ of the burst is
determined by the the radius $R$ where the optical depth of the
material ahead of the shock drops to $\sim1$, i.e. $\delta t\sim
R(\tau=1)/c$ (If the stellar wind surrounding the progenitor were
optically thin everywhere, the shock would break out from the SN
progenitor stellar envelope and the variability time would be
about $R_{\star}/c$, where $R_{\star}$ is the stellar radius.)

\subsection{Monte Carlo simulation of photon ``acceleration''}
\begin{figure*}
\centering 
\includegraphics[width=10cm]{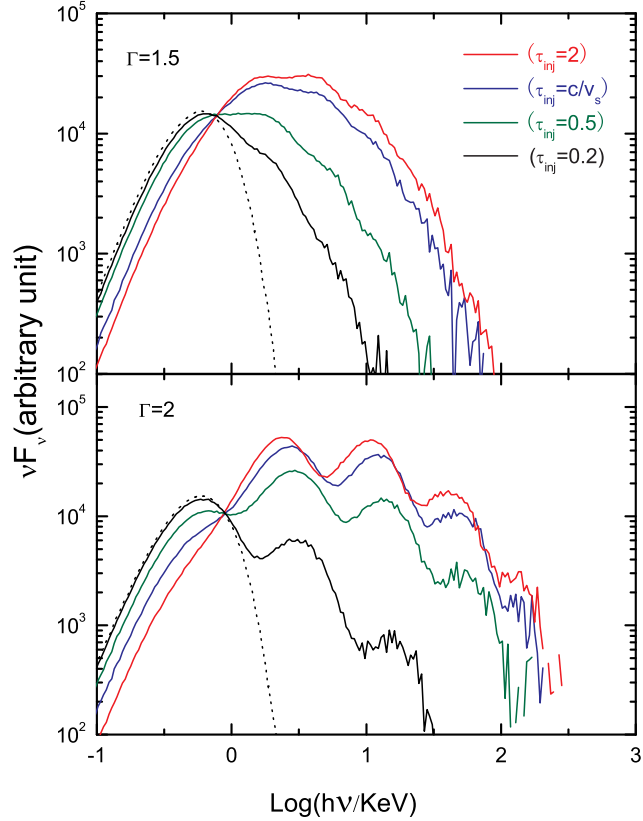}
\caption{\small  The time-integrated energy distribution of the
escaping photons. $10^6$ photons, with a black body distribution
at $k_B T_{th}=0.15{\rm KeV}$ (black dotted line), are injected at
four different times, corresponding to optical depths of the
pre-shocked medium of $\tau_{\rm inj}=2$, $c/v_s$, 0.5 and 0.2.
The ejecta Lorentz factors are $\Gamma=1.5$ and $\Gamma=2$ for
upper panel and lower panel respectively. Note, that the ``humps''
seen in the spectra are an artifact of the one-dimensional
simulation, and are expected to be smoothed out in reality. See
\cite{Wang et al. 2007} for more details.}
\end{figure*}

In order to understand the photon ``acceleration'' mechanism in
the time-dependent case, we carried out a Monte Carlo simulation
of repeated Compton scattering during shock breakout from a dense
stellar wind \cite{Wang et al. 2007}. {We approximate the
hydrodynamics of the problem as follows. We consider the mildly
relativistic ejected shell to act as a piston with a
time-independent bulk Lorentz factor $\Gamma$ and an infinite
optical depth $\tau_{ej}$. Since the stellar wind swept up by the
ejecta is not sufficient to decelerate it, the constant velocity
of the interface is justified. This``piston" drives a shock into
the surrounding medium, where the density profile is assumed to
follow $n \propto R^{-2}$. We assume the shock width to be
infinitesimal. There are three distinct regions in this picture:
The moving piston, the shocked medium and the pre-shocked medium.
The shocked medium is considered to form a homogeneous shell, and
the electrons are regarded as cold, with a bulk velocity same as
that of the ejecta. The velocity of the shock front can then be
obtained consistently from the shock jump condition.   We further
simplify the problem by considering a one-dimensional situation,
where the motion of the photons is confined to one dimension,
forward or backward relative to the shock expansion direction. The
photons are injected at the shock front and would then be
repeatedly scattered among these three components until they
escape out to a sufficiently further region. Our simulation takes
into account the proper possibility of photon scattering by each
of the three components (with the appropriate photon energy gain
or loss). The photon scattering probability by the unshocked wind
and shocked wind is determined by the optical depth of each
component, while the piston is regarded as a mirror since it has a
larger optical depth much larger than unity.}

We study the spectra of escaping photons, which result from the
``injection'' of photons at the shock front at various radii,
corresponding to various values of $\tau$ at the injection time,
denoted by $\tau_{\rm inj}$. The photons are treated as ``test
particles'', and their scattering history is followed as the shock
expands and $\tau$ decreases. The resulting time-integrated $\nu
F_\nu$ spectra are given in Fig.~1 for two values of $\Gamma$,
$\Gamma=2$ and $\Gamma=1.5$. The red, blue, olive and black curves
describe the spectrum of escaping photons resulting from the
injection of a thermal distribution of photons at $\tau_{\rm
inj}=2$, $c/v_s$, $0.5$ and $0.2$ respectively. The injected
photons are assumed to have a thermal spectrum with $T=0.15$~keV.
The "humps" seen in the simulated spectra correspond to different
orders of Compton scattering, with two nearby humps separated by
$\sim\Gamma^2$. These humps appear because we do not consider in
our one-dimensional simulation the angular distribution of
scattered photons. In reality, these humps are expected to be
smoothed out. Fig.~1 demonstrates that a significant fraction the
thermal photons that are injected at optical depth $\tau_{\rm
inj}\sim1$, where the photons are expected to escape the shock,
may be "accelerated" to high energy. The resulting non-thermal
component may carry a significant fraction of the shell energy,
and its luminosity could exceed that of the thermal component,
depending on the value of $\Gamma$. As expected, the spectrum
depends on $\Gamma$ and on $\tau_{\rm inj}$, with harder spectra
obtained for larger values of $\Gamma$ and $\tau_{\rm inj}$. Note
that the spectra are steeper (softer) at high energies, due to the
longer time required for acceleration to higher energies, which
implies a significant decrease in $\tau$ during the acceleration
process.

\section{Shock breakout origin of the X-ray outburst XRO 080109}

Although the thermal x-ray  component is not identified in  XRO
080109, there are a few facts that favor the shock breakout
interpretation: 1) smooth light curve--consistent with shock
breakout prediction; 2) low total radiation energy of $10^{46}
{\rm erg}$--close to the theory prediction for a normal type Ib/c
SN; 3) early (few days) decline in the optical emission from SN,
probably optical emission from the adiabatically cooling ejecta
heated by shock breakout (\cite{Waxman et al. 2007}).

The non-thermal spectrum of XRO 080109  can be due to the
bulk-comptonization of the thermal shock breakout photons. Then,
the peak of the thermal component should be not larger than 0.3KeV
(the low threshold energy of XRT) so that the thermal component
was not detected by XRT. Since the Compton $Y$ parameter is
typically of the order of unity, the energy in non-thermal
emission should be roughly comparable to that in thermal energy.
From these simple arguments, we now derive the constraint on the
shock parameters, following \cite{Waxman et al. 2007}.

For a strong radiation-dominated shock, the radiation pressure
$p=f (\Gamma\beta)^2 \rho c^2$, where $\rho$ is the mass density
of the pre-shocked density and $f\simeq0.8$ for trans-relativistic
shock (\cite{Waxman et al. 2007}). The post shock temperature
$T_d$ is related to the the postshock pressure by $a T_d^4=3p$,
and the observed temperature is $T=\gamma_d T_d$, where $\gamma_d$
is the Lorentz factor of the downstream flow.  The temperature of
the shock breakout thermal emission $T$ is related to the velocity
of the shock and photosphere radius by
\begin{equation}
a T^4= 3f (\Gamma\beta)^2 \rho c^2 \gamma_d^4
\simeq(\Gamma\beta)^6{ \rho c^2}=(\Gamma\beta)^6 c^2
(\frac{1}{\kappa R_{ph}}),
\end{equation}
where the optical depth $\tau(R)=\kappa R \rho$ with
$\tau(R_{br})=\tau(R_{ph})=1$ at the  breakout radius $R_{br}$ for
trans-relativistic shock and $\kappa$ is the Thomson opacity.

Assuming  the energy in the thermal component is comparable to
that in the non-thermal one (which is likely for a
trans-relativistic shock), i.e. $E_{th}\sim 2\times10^{46} {\rm
 erg}$, we can obtain, from Eq.(2) in ref.\cite{Waxman et al. 2007},
 \begin{equation}
(\Gamma\beta)^2\simeq\frac{E_{th}}{0.5\times 4\pi R_{ph}^3 \rho
c^2}=\frac{E_{th} \kappa}{2\pi R_{ph}^2 c^2 }.
\end{equation}

Combining Eqs.(3) and (4), we obtain
\begin{equation}
\Gamma\beta=1.1 (\frac{T}{0.1KeV})^{2/7}
(\frac{E_{th}}{2\times10^{46}{\rm erg}})^{1/28}(\frac{\kappa}{0.2
g^{-1} cm^2})^{3/28}
\end{equation}
and
\begin{equation}
R_{br}=7\times10^{11}
(\frac{T}{0.1KeV})^{-4/7}(\frac{E_{th}}{2\times10^{46}{\rm
erg}})^{3/7} (\frac{\kappa}{0.2 g^{-1} cm^2})^{2/7}{\rm cm},
\end{equation}
where $\kappa\simeq 0.2 {\rm g^{-1} cm^{2}}$ is the Thompson
opacity for ionized He wind.

Requiring $T\leq{\rm 0.1 KeV}$ to account for the non-detection of
the thermal peak, we derive the constraints on the shock velocity
$\Gamma\beta $ and breakout radius $R_{br}$, i.e.
\begin{equation}
\Gamma\beta\leq 1.1 (\frac{E_{th}}{2\times10^{46}{\rm
erg}})^{1/28}(\frac{\kappa}{0.2 g^{-1} cm^2})^{3/28},
\end{equation}
\begin{equation}
R_{br}\geq 7\times10^{11} (\frac{E_{th}}{2\times10^{46}{\rm
erg}})^{3/7} (\frac{\kappa}{0.2 g^{-1} cm^2})^{2/7}{\rm cm} .
\end{equation}
We note that there is a misuse of the relation $L_{rad}=4\pi R^2
\sigma T^4$ to derive the shock breakout radius in some papers,
where $T$ is the radiation temperature. This is because  the light
travel time (~$R/c$) will lengthen the breakout emission duration
significantly and render the observed radiation luminosity lower
than $4\pi R^2 \sigma T^4$ significantly (see also \cite{Matzner &
Mckee 1999}). So one would get a much smaller radius for shock
breakout if this incorrect  relation is used.

The rise time of the x-ray transient is about 60  s
(\cite{Soderberg et al. 2008}, implying a radius for the emitting
region
\begin{equation}
R_{ph}\simeq c\delta t = 2\times10^{12}  {\rm cm},
\end{equation}
which is consistent with the above inferred breakout radius. Since
this radius is larger than that of the WR progenitor, the shock
must break out from the surrounding stellar wind.  From $\tau=1$
at $R_{br}$, we derive $\dot{M}=4\pi v_w
R_{br}/\kappa\geq7\times10^{-5} {\rm M_\odot yr^{-1}}
({\kappa}/{0.2 g^{-1} cm^2})^{-1}$ for $v_w=1000 {\rm Km s^{-1}}$.

So we need a trans-relativistic shock with $\Gamma\beta\leq1$ at
the time when shock is breaking out. This can be achieved for a
normal type Ib/s SN explosion, according to the calculation of
\cite{Matzner & Mckee 1999}. Substituting the inferred ejecta
kinetic energy ($E_K=2-4\times10^{51}{\rm erg}$) and mass
($M_{ej}=3-5 M_\odot$) the of  SN 2008D and a typical Wolf-Rayet
star radius of $R=10^{11}{\rm cm}$ into Eq.(32) of \cite{Matzner &
Mckee 1999}, the maximum velocity that the radiation-dominated
shock can reach is trans-relativistic.

With  a velocity $0.5\leq\Gamma\beta\leq1$ , the energy
amplification factor for one Compton scattering is
\begin{equation}
A=\Gamma^2(1+\beta^2/3)=1.5-2.3
\end{equation}
which implies an equivalent Compton parameter
$Y=A(1-e^{-\tau})\simeq 1$, consistent with our earlier
assumption.

\section{Discussions}
Chevalier \& Fransson \cite{Chevalier & Fransson 2008} questioned
the bulk Comptonization mechanism for XRO080109/SN2008D by arguing
that "the breakout radiation is capable of accelerating the matter
ahead of the shock front so that the formation of the gas
dominated shock is delayed". This is not true because as long as
there is a converging flow (in our case the supernova ejecta
moving relative to the pre-shock matter), the repeated scatterings
will occur and the bulk Comptonization will work. So the formation
of an viscous shock is not a necessity for the bulk Comptonization
process. Also, for a type Ib/c supernova shock that reach a
mildly-relativistic velocity, the radiation is actually incapable
of accelerating the matter ahead of the shock front to such a
mildly-relativistic velocity, according to our calculation given
below. The radiative acceleration is
\begin{equation}
g_R=\frac{\kappa L_{rad}}{4\pi R^2 c}=5\times10^7
(\frac{\kappa}{0.2 g^{-1} cm^2})(\frac{L_{rad}}{10^{42} {\rm erg
s^{-1}}}) R_{12}^{-2} \,\, {\rm cm \, s^{-2}}.
\end{equation}
So the maximum velocity that the matter ahead of the shock front
be accelerated to is
\begin{equation}
v_m=g_R \Delta t=5\times10^9(\frac{\kappa}{0.2 g^{-1}
cm^2})(\frac{L_{rad}}{10^{42} {\rm erg s^{-1}}}) R_{12}^{-2}
(\frac{\Delta t}{100 \rm s}) \, {\rm cm s^{-1}},
\end{equation}
which is below the mildly-relativistic velocity of the shock
front.

\begin{theacknowledgments}
{We would like to thank Alicia Soderberg for fruitful
collaboration. XYW also thank R. Chevalier for useful discussions
at the 37th Cospar meeting. This work is supported by the National
Natural Science Foundation of China under grants 10221001 and 973
projects under grants 2009CB824800, and the Foundation for the
Authors of National Excellent Doctoral Dissertations of China.}
\end{theacknowledgments}


\begin{thebibliography}{99}



\bibitem{Colgate 1974}
Colgate, S. A.  {\em ApJ}, 187: 333, 1974
\bibitem{Klein & Chevalier 1978}
Klein, R. I. \& Chevalier, R. A., {\em ApJ}, 223: L109, 1978
\bibitem{Ensman & Burrows 1992}
Ensman, L. \& Burrows, A.,  {\em ApJ}, 393: 742, 1992
\bibitem{Matzner & Mckee 1999}
Matzner, C.~D., \& McKee, C.~F., {\em ApJ}, 510: 379, 1999
\bibitem{Campana et al. 2006}
Campana, S. et al.  {\em Nature}, 442: 1008, 2006
\bibitem{Waxman et al. 2007}
Waxman, E., \Meszaros, P. \& Campana, S. {\em ApJ}, 667:351, 2007


\bibitem{Wang et al. 2007}
Wang, X. Y., Li, Z., Waxman, P. and \Meszaros, P., \emph{ApJ},
664:1026, 2007
\bibitem{Soderberg et al. 2008}
Soderberg, A. et al., {\em Nature}, 453:469, 2008
\bibitem{Xu et al. 2008}
Xu, D., Zou, Y. C. and Fan, Y. Z., preprint (arXiv:0801.4325),
2008
\bibitem{Li 2008}
Li, L. X., {\em MNRAS}, submitted (arXiv:0803.0079), 2008
\bibitem{Mazzali et al. 2008}
Mazzali, P. et al. {\em Science}, accepted (arXiv:0807.1695), 2008
\bibitem{Blandford & Payne 1981}
Blandford, R. D. \& Payne, D. G.,  {\em MNRAS}, 194: 1041, 1981
\bibitem{Gruzinov & Meszaros 2000}
Gruzinov, A. \& \Meszaros,  {\em ApJ}, 539: L21, 2000

\bibitem{Thompson 1994}
Thompson, C. 1994, {\em MNRAS}, 270: 480

\bibitem{Chevalier & Fransson 2008}
Chevalier, R. and Fransson C., {\em ApJ}, submitted
(arXiv:0806.0371), 2008

\end{thebibliography}
\end{document}